\documentclass[fleqn,usenatbib]{mnras}

\usepackage{newtxtext,newtxmath}

\usepackage[T1]{fontenc}
\usepackage{ae,aecompl}
\usepackage[normalem]{ulem}

\def\draftversion{0} 
\usepackage{amssymb,latexsym,graphicx,natbib,eufrak,times,amsmath,xspace,ifthen}

\ifthenelse{ \draftversion > 0 }{
} {
  
}

\ifthenelse{\equal{\draftversion}{1}}{
  \usepackage{xcolor}
  \newcommand{\sep}[1]{\par\begin{color}[rgb]{0,0.4,0}\begin{center}\hrule\end{center}\end{color}\par} 
  \newcommand{\todo}[1]{\begin{color}{red}\ \ifthenelse{\equal{#1}{}} {$\bullet\bullet\bullet$} {$\bullet$\ #1 $\bullet$}\end{color}} 
  \newcommand{\idea}[1]{\begin{color}[rgb]{0,0.4,0}\textit{#1}\end{color}} 
  \newcommand{\sk}[1]{\begin{color}[rgb]{0.6,0,0.6}#1\end{color}} 
  \newcommand{\toc}{\par\begin{color}[rgb]{0.6,0,0.6}\begin{center}\hrule\vspace{0.5mm}\begingroup\small\let\cleardoublepage\relax\let\clearpage\relax\mytoc\endgroup\vspace{0.5mm}\hrule\end{center}\end{color}\par} 

  \newcommand{\OA}[1]{\begin{color}{red}\textbf{OA: }#1\end{color}\xspace}
  \newcommand{\EA}[1]{\begin{color}{red}\textbf{EA: }#1\end{color}\xspace}
  \newcommand{\oa}[1]{{\bf \textcolor{red}{OA:#1}}}
  }{
  \newsavebox{\trashcan}

  \newcommand{\sep}[1]{}
  \newcommand{\todo}[1]{}
  \newcommand{\idea}[1]{}
  \newcommand{\sk}[1]{}
  \newcommand{\toc}{}

  \newcommand{\OA}[1]{}
  \newcommand{\EA}[1]{}
  \newcommand{\oa}[1]{}
  }
 \makeatletter\newcommand\mytoc{\@starttoc{toc}}\makeatother 
\long\def\symbolfootnote[#1]#2{\begingroup%
\def\thefootnote{\fnsymbol{footnote}}\footnote[#1]{#2}\endgroup} 

\newcommand{\eqn}[2][]{Equation#1~\ref{eq:#2}} 
\newcommand{\fig}[2][]{Figure#1~\ref{fig:#2}}

\newcommand{\sect}[2][]{Section#1~\ref{sec:#2}}

\newcommand{\bb}[1]{\ifmmode \mbox{\boldmath $ #1$} \else  \mbox{\boldmath $#1$} \fi}

\newcommand{\U}[1]{\ensuremath{\mathrm{~#1}}}     

\newcommand{\yr}{\U{yr}}
\newcommand{\Myr}{\U{Myr}}          
\newcommand{\Gyr}{\U{Gyr}}          
\newcommand{\pc}{\U{pc}}
\newcommand{\kpc}{\U{kpc}}
          
\newcommand{\Msun}{\U{M}_{\odot}\xspace}   \newcommand{\msun}{\Msun}
\newcommand{\Msunyr}{\Msun\yr^{-1}} 
   
\newcommand{\cc}{\U{cm^{-3}}}
\newcommand{\K}{\U{K}}
    
\newcommand{\kms}{\U{km\ s^{-1}}}
\newcommand{\erg}{\U{erg}}

\newcommand{\hi}{H{\sc i} }                       

\newcommand{\kick}{{\tt runaways}\xspace}
\newcommand{\nokick}{{\tt no\,runaways}\xspace}

\newcommand{\ramses}{{\small RAMSES}\xspace}

\newcommand{\agora}{{\small AGORA}\xspace}

\newcommand{\lund}{Department of Astronomy and Theoretical Physics, Lund Observatory, Box 43, SE-221 00 Lund, Sweden}

\title[Increased outflows with runaways]{How runaway stars boost galactic outflows}

\author[E. P. Andersson et al.]{
Eric P. Andersson$^{1}$\thanks{E-mail: eric@astro.lu.se},
Oscar Agertz$^{1}$ and
Florent Renaud$^{1}$
\\
$^{1}$ \lund \\
}

\date{Accepted XXX. Received YYY; in original form ZZZ}

\pubyear{2020}

\begin{document}
\label{firstpage}
\pagerange{\pageref{firstpage}--\pageref{lastpage}}
\maketitle

\begin{abstract}
Roughly ten per cent of OB stars are kicked out of their natal clusters before ending their life as supernovae. These so-called runaway stars can travel hundreds of parsecs into the low-density interstellar medium, where momentum and energy from stellar feedback is efficiently deposited. In this work, we explore how this mechanism affects large-scale properties of the galaxy, such as outflows. To do so we use a new model which treats OB stars and their associated feedback processes on a star-by-star basis. With this model we compare two hydrodynamical simulations of Milky Way-like galaxies, one where we include runaways, and one where we ignore them. Including runaway stars leads to twice as many supernovae explosions in regions with gas densities ranging from $10^{-5}\,\cc$ to $10^{-3}\,\cc$. This results in more efficient heating of the inter-arm regions, and drives strong galactic winds with mass loading factors boosted by up to one order of magnitude. These outflows produce a more massive and extended multi-phase circumgalactic medium, as well as a population of dense clouds in the halo. Conversely, since less energy and momentum is released in the dense star forming regions, the cold phase of the interstellar medium is less disturbed by feedback effects.

\end{abstract}






\section{Introduction}\label{sec:intro}
$\Lambda$-cold dark matter has been highly successful in explaining and predicting a variety of observed properties, such as large scale structure, halo clustering and galaxy scaling relations \citep{Eisenstein+2005,Springel+2005b,Viel+2008,Reid+2010,Klypin+2011,Komatsu+2011,Somerville+2015}. Nonetheless, this model has also encountered challenges related to the `baryon cycle', i.e. how galaxies accrete and expel their gas. Galaxy formation is an inefficient process, with at most $\sim1/3$ of the cosmological baryon fraction being turned into stars in galaxies as massive as the Milky Way, and significantly less in dwarf galaxies \citep[e.g.][]{Behroozi2019}. This notion has been notoriously difficult to predict by numerical simulations of galaxy formation, which historically have suffered from excessive gas cooling and loss of angular momentum, leading to simulated galaxies with little in common with observations \citep[for a review on this topic, see][]{Naab&Ostriker2017}. 

Galactic outflows, driven by stellar feedback \citep[][]{DekelSilk86} and active galactic nuclei \citep[]{Benson2003}, are commonly believed to be instrumental for solving these problems \citep[][]{Somerville+2015}. In the past decades, many studies have explored ways of numerically capturing the impact of stellar feedback processes on the interstellar medium (ISM), e.g. effects from supernovae (SNe) explosions, stellar winds, radiation from young stars \citep[see, e.g.,][]{Katz1992,Stinson+2006,Agertz+2013,Keller+2014,Simpson+2015,FIRE2}. This effort has made it possible to model galactic outflows in a cosmological context as an emergent property of clustered star formation, with simulations now matching a range of observed galaxy properties \citep[e.g.][]{Hopkins2014,AgertzKravtsov2015,AgertzKravtsov2016}, as well as properties of the turbulent ISM and giant molecular clouds (GMCs; \citealt{Grisdale2017,Grisdale2018,Grisdale2019}). 

Adding to the success of stellar feedback models, simulations of entire galaxies has improved significantly in the recent years with parsec, or even sub-parsec spatial resolution and mass resolution reaching a few solar masses enabling models to capture the dense, cold ISM \citep[see, e.g.,][]{Renaud2013,Rosdahl+2017,Wheeler+2018,AgertzEDGE2019}. State-of-the-art simulations can now resolve most of the cooling radii of individual SN explosions which allows them to capture the build-up of momentum during the Sedov-Taylor phase which is crucial for robustly modelling SNe feedback \citep[][]{Kim&Ostriker2015,Martizzi2015}. Furthermore, recent numerical studies have reached high enough resolution for individual stars to be modelled in galaxy scale simulations \citep{Hu+2017,Wheeler+2018,Su+2018,Emerick2019,Lahen2019}. We stress however that despite the improvements in galaxy modelling in the past decade, (subgrid) stellar feedback still operates at the resolution level, with its coupling to the ISM not yet being fully understood \citep[][]{Ohlin+2019}. 

While the actual process of star formation is not yet captured in galactic context, these simulations allow for star particles to represent individual stars, all sampled from an initial mass function (IMF) with subsequent feedback processes emerging on a star-by-star basis rather than from macro particles representing entire stellar populations. However, while galaxy simulations have started to resolve the stellar component star-by-star, they are still far from treating the stars \emph{collisionally}. While a collisionless approximation is valid on galactic scales, star-star interactions drive the internal dynamics of dense star clusters. As a result, some fraction of OB stars are kicked out of their natal star clusters with velocities large enough to move them out of the dense star forming gas before exploding as SNe \citep[e.g.,][]{Gies&Bolton1986,Gies1987,Stone1991,Hoogerwerf+2000,Schilbach&Roser2008,Jilinski2010,Silva&Napiwotzki2011,Appelaniz+2018}. Simulation of clusters \citep[e.g.,][]{Oh&Kroupa2016} reveal that such velocity kicks originates from gravitational interactions \citep[][]{Poveda+1967}, including the disruption of binaries through SNe \citep{Blaauw1961}. None of these effects can be taken into account self-consistently without collisional dynamics, which currently is computationally infeasible in galaxy scale simulations. 

Massive runaway stars have been suspected to impact galaxy evolution through efficient injection of energy into low density gas \citep[][]{Naab&Ostriker2017}. Coupling energy from SNe to gas depends both on the phase of the gas \citep{Katz1992,Ceverino&Klypin2009}, as well as on the structure of the ISM \citep{Kim&Ostriker2015,Martizzi+2015,Walch&Naab2015,Iffrig&Hennebelle2015,Ohlin+2019}. Because runaway stars move away from their natal regions, SN energy from these objects ought to couple differently compared to the stars which remains in closer proximity to the dense star forming gas. Typically explosions in low density gas generate a hot bubble which expands and leaves the galaxy as an outflow \citep{Korpi+1999,deAvillez2000,deAvillez&Breitschwerdt2004,Joung&MacLow2006}. A similar effect has also been studied by comparing feedback from SNe positioned at density peaks compared to random positioning \citep[see e.g.][]{Korpi+1999,deAvillez2000,deAvillez&Breitschwerdt2004,deAvillez&Breitschwerdt2005,Joung&MacLow2006,Walch+2015,Martizzi+2016,Girichidis+2016,Gatto+2017}. Although these experiments are idealised, targeting only a small patch of the galaxy, they demonstrate that feedback from randomly located SN tends to drive stronger outflows compared to explosions in density peaks. This potentially impact the way the host galaxy forms and evolves across cosmic time. 

In this paper we present a new model for treating individual stars, focusing on the effect of runaways. The model has been applied to simulations of an isolated disc galaxy with properties similar to the Milky Way. We describe how we model individual stars as well as our treatment of runaway stars in \sect{star-formation_and_feedback}. The numerical setup and results are described in \sect{isodisc_sim_setup} and \ref{sec:results} respectively, while we discuss their implications in \sect{discussion}. Finally we summarise in \sect{conclusions}.

\section{Star formation and stellar feedback}\label{sec:star-formation_and_feedback}
In order to probe the effects of runaway stars, we implement a new recipe for treating massive stars as individual particles which evolves during the simulation. In this section we describe the method used for sampling stars from an IMF, how feedback via winds and SNe is treated and how it chemically enriches the ISM.

\subsection{Sampling the initial mass function}\label{subsec:sampling_imf}
Treating massive stars as individual particles requires a method to sample stellar masses. To do this efficiently we employ the method by \citet{Sormani+2017}. In a mass bin indexed by $i$, a number of stars $n_i$ is determined through realisations of a Poisson distribution given by 
\begin{equation}\label{eq:imf_poisson_sampling}
\mathcal{P}_i(n_i)=\frac{\lambda_i^{n_i}}{n_i!}{\rm e}^{-\lambda_i},
\end{equation}
where the mean of the distribution, $\lambda_i$, is given for each bin by
\begin{equation}\label{eq:imf_poisson_parameter}
\lambda_i=f_i\frac{M}{m_i},
\end{equation}
where for bin $i$, in which stars have an average mass of $m_i$, there is a fraction of mass $f_i$ out of the total mass $M$, available for star formation. The total stellar mass generated in $N$ bins is then
\begin{equation}\label{eq:imf_total_mass}
M_{\rm sampled} = \sum_{i=1}^N n_im_i,
\end{equation}
which is approximately equal to the mass $M$, due to the stochastic sampling, but converges towards $M$ for large numbers of stars. There are two properties of this method which makes it ideal for our simulations. Firstly, the computational expense is determined by the number of adopted mass bins, since the method only samples one random number for each bin and not for every star. Secondly, the bin-sizes can be chosen arbitrarily, which allows for stars below a certain mass threshold to be grouped in a single bin, while stars above this mass are sampled at high mass resolution. 

A problem with this method is the non-zero probability of sampling a population of stars with mass larger than the available gas mass. To avoid this, we sample the IMF starting from the low-mass end until either reaching the most massive stars defined by the mass bins or until the available gas is depleted. In the first case we return the extra mass to the gas parcel where it was initially collected, while in the second case we remove the most massive stars such that the stellar mass formed does not exceed the initial gas mass. This results in an IMF which is slightly more bottom heavy than the one originally targeted, which in principle could lead to weaken stellar feedback. Nonetheless, this effect is negligible compared to the uncertainties in the models for stellar feedback. 

Our model uses of the three part IMF from \citet{Kroupa2001}, where the number of stars of mass $m$ is given by 
\begin{equation}\label{eq:kroupa_imf}
\xi(m)=AC_jm^{-\alpha_j} \quad {\rm for} \quad m_j \leq m < m_{j+1},
\end{equation}
where $A$ is a normalisation factor and $C_j$ are constants that ensures continuity in the intersections given by $C_1=1$, $C_2=C_1m_2^{\alpha_2-\alpha_1}$ and $C_3=C_2m_3^{\alpha_3-\alpha_2}$. The three different parts are split into mass regimes given by $m_1=0.01\msun$, $m_2=0.08\msun$, $m_3=0.5\msun$ and $m_4=M_{\rm max}$, with slopes $\alpha_1=0.3$, $\alpha_2=1.3$ and $\alpha_3=2.3$. The normalisation is determined for stars with masses between $M_{\rm min}=0.01\msun$ and $M_{\rm max}=120\msun$. 

In this work we sample stars from the aforementioned IMF at each star formation event. We divide stars into high mass stars (HMS) and low mass stars (LMS). The HMS are defined to have stellar masses larger than $8\msun$, sampled up to $40\msun$ using 100 equisized bins (giving a mass resolution of $\sim\!0.3\msun$). We do not include stars more massive than $40\Msun$ in our model. Such stars are both exceedingly rare and their short lifetimes means that the distance traveled by very massive runaway stars is short. However, these are extremely luminous important sources of feedback, especially in regards to stellar winds. Our model for stellar winds (described in detail in \sect{stellar_feedback}) has a strong scaling with stellar mass, and including these most massive stars can result in too strong early feedback. Furthermore, we assume core-collapse SNe for all HMS, which is unrealistic for too massive stars. In the code, all HMS are treated individual particles. Since we are interested in the effect of feedback from massive ($>8\msun$) runaway stars, the rest of the IMF, defining the LMS, are lumped together into a single macro particle to reduce the computational cost. 

\subsection{Runaway stars}\label{sec:birth_kicks}
In the simulations, particles are treated in a collisionless manner. To simulate runaway stars, we give velocity kicks to all particles representing individual stars (i.e. $\geq8\msun$) at their formation. The collisional effects leading to kicks operate on spatial scales many orders of magnitudes below the gravitational softening length of our simulations, warranting a `sub-grid' approach. 

The method assumes kicks distributed isotropically with velocities, $v$, sampled from a power-law distribution,
\begin{equation}\label{eq:velocity_distribution}
    f_{v}\propto v^{-\beta},
\end{equation}
where the $\beta$ is a free parameter in our model. The value of $\beta$ depends on several factors. Dynamical scattering events which generate kicks typically involve the interaction between a hard binary and a third star. The frequency of such encounters will therefore not only depend on the relaxation time of the cluster, but must also be sensitive to the binary fraction. Moreover, as mentioned earlier, the kicks can also originate from the disruption of binary system caused by the SNe of one of the two components. This corroborates that binary fraction is a important parameter of the distribution of velocities. Furthermore, these processes implies a delay between the formation of a star and the time when it gets a kick. Because of the complexity of this problem, estimates of $\beta$ typically demands the use of N-body simulations \citep[see, e.g.,][]{Eldridge+2011,Perets+2012,Oh&Kroupa2016,Renzo+2019}, but estimates from observations also exists \citep[e.g.,][]{Hoogerwerf+2000,Hoogerwerf+2001}. Furthermore, \cite{Banerjee+2012} found that the velocity distribution shows some dependence on the mass of the runaway stars, with more massive stars reaching higher velocities \citep[in agreement with][]{Oh&Kroupa2016}. 

In this work, we choose a value of $\beta=1.8$ and normalise the distribution for values between $3\kms$ and $385\kms$, which we apply to all HMS particles at formation without adding any time delay. This leads to $\sim\!14\%$ stars with velocities $>\!30\kms$. This choice is motivated by model MS3UQ\_SP in \citet{Oh&Kroupa2016}, corresponding to runaways from a clusters with a mass of $10^{3.5}\Msun$ and half mass radii of $0.3\pc$. Early observations indicated values for the runaway fraction of $30\%$ \citep{Stone1991}, however this is large compared to more recent work. As noted by \citealt{Appelaniz+2018} (see also \citealt{Silva&Napiwotzki2011}), the results of \citeauthor{Stone1991} overestimates this fraction. \citeauthor{Appelaniz+2018} found observational evidence that $10\!-\!12\%$ of O stars and a few percent of B stars are so called runaway stars with significant peculiar velocities $(>\!30\kms)$ with respect to their natal environment \citep[in agreement with models by][]{Eldridge+2011,Renzo+2019}. Furthermore, the sampled velocities applied to the stars will change through gravitational forces acting on the stars throughout their lifetime. This change depends on the local gravitational field and the entire galactic potential for stars with long enough life times. This sensitivity to environment, together with the difficulty of comparing the observed population of runaway stars to the un-evolved velocity distribution, make a simple universal model (e.g. \eqn{velocity_distribution}) uncertain. The reader should be cautious of this and note that the results we present could overestimate the effect of runaway stars. 

As a first approximation, one can naively compute the distance stars travel before exploding as SNe by multiplying their mean velocity with the average main-sequence life time in the appropriate mass range. Using the velocity distribution given by \eqn{velocity_distribution} we find that stars in the mass range $8\msun$ to $40\msun$ with $Z = Z_\odot$\footnote{in this work we adopt $Z_\odot=0.02$.} travels roughly $350\pc$ before exploding as supernova. This simple model does not take into account deceleration from the gravitational field of the cluster from which it escapes, and is therefore a upper limit on the travel distance. Nonetheless it is of the same order of magnitude as more detailed studies \citep[see e.g.][]{Eldridge+2011,Renzo+2019}. The distance is significant, being an order of magnitude larger than the average size of star forming molecular clouds in the Milky Way \citep[e.g.][]{Heyer2009}, and of the same order of magnitude as the scale height of the cold gas disc.

\subsection{Stellar feedback}\label{sec:stellar_feedback}
HMS particles are considered as individually evolving stars and we consider mass loss, enrichment, momentum- and energy-injection from fast winds and core-collapse SNe. In massive stars ($\geq8\msun$), radiation pressure is significant enough to push away their outer envelopes, giving rise to strong stellar winds for their entire main-sequence evolution \citep[see, e.g.,][]{Willis&Germany1987, Cassinelli&Lamers1987}. For this work we employ a model with a mass-loss rate based on \citet{Dale&Bonnell2008} given by
\begin{equation}
    \dot{M} = 10^{-5}\left ( \frac{M_{\mathrm{b}}}{30\,\mathrm{M}_{\odot}}\right )^4\,Z^{\gamma}\,\mathrm{M}_{\odot}\,\mathrm{yr}^{-1},
\end{equation}
where $M_{\mathrm{b}}$ is the birth mass of the star. The scaling to the metallicity\footnote{We track  iron  (Fe)  and  oxygen  (O)  abundances  separately, and advect them as passive scalars. To construct a total metal mass $M_{Z}$ to use for feedback and cooling,via the metallicity $Z=M_{Z}/M_{\rm gas}$, we adopt $M_{Z}=2.09M_{\rm O}+1.06M_{\rm Fe}$ according to the mixture of $\alpha$ (C, N, O, Ne, Mg, Si, S) and iron (Fe, Ni) group elements relevant for the sun \citep{Asplund2009}.} was added to the mass-loss rate to account for the metallicity dependence of the photon coupling to the stellar envelope, which drives the wind. The metallicity exponent $\gamma$ has been shown to range between 0.5 and 0.8, \citep{Kudritzki+1987,Vink+2001,Mokeim+2007,Vink2011} and in this work we adopt $\gamma = 0.5$. The velocity of the fast winds typically ranges between $1000\kms$ and $3000\kms$ in the literature \citep[see, e.g.,][]{Leitherer1992, Lamers&Cassinelli1999}. We use a value $v_{\mathrm{w}}=1000\kms$ for our model, and inject the wind into the surrounding gas as a continuous source of momentum during the star's main sequence life time.

HMS stars explode as core-collapse SNe at the end of their main sequence. The main sequence lifetime, $t_{\mathrm{MS}}$, for a star given its mass and metallicity $Z$ (here expressed as the total mass fraction in elements heavier than He). To determine $t_{\rm MS}$ our model uses a stellar age-mass-metallicity fit by \citet{Raiteri+1996}, who found 
\begin{equation}\label{eq:main_sequence_lifetime}
    \log t_{\mathrm{MS}} = a_0(Z) + a_1(Z)\log m + a_2(Z)(\log m)^2, 
\end{equation}
where the coefficients are given by 
\begin{equation}
\begin{array}{l}
    a_0(Z) = 10.13 + 0.07547\log Z - 0.008084(\log Z)^2, \\
    a_1(Z) = - 4.424 - 0.7939\log Z - 0.1187(\log Z)^2, \\
    a_2(Z) = 1.262 + 0.3385\log Z + 0.05417(\log Z)^2.
\end{array}
\end{equation}
Note that this gives very similar main sequence lifetimes compared to the single-star evolution (SSE) formulae by \citet{Hurley+2000}. When a star explodes we deposit $10^{51}\erg$ of energy into the gas at the location of the star particle. If the SNe cooling radius is not resolved, hence the momentum built up during the Sedov-Taylor phase is not captured self-consistently, we explicitly inject the momentum from this phase following the scheme by \citet[see also \citealt{Martizzi2015}]{Kim&Ostriker2015}. Our implementation is identical to that of \citet[][]{Agertz2015} and \citet[][]{Rhodin2019} and we refer the reader to these works for more details.

For LMS particles we consider mass loss, enrichment, momentum- and energy-injection for type Ia SNe and asymptotic giant branch (AGB) winds. Details of the scheme can be found in \citet{Agertz+2013}, with the adopted Kroupa IMF (\eqn{kroupa_imf}) truncated at at the LMS upper mass limit ($8\msun$).

\section{Numerical setup}\label{sec:isodisc_sim_setup}
We have implemented the method described above in the $N$-body + Adaptive Mesh Refinement (AMR) code \ramses \citep{Teyssier2002}. \ramses treats stars and dark matter through collisionless dynamics using the particle-mesh method \citep{Hockney&Eastwood1981,Klypin&Shandarin1983}, with the gravitational potential calculated by solving the Poisson equation using the multi-grid method \citep{GuilletTeyssier2011} for all refinement levels. The fluid dynamics of the gas is calculated using a second-order unsplit Godunov method with an ideal mono-atomic gas with adiabatic index $\gamma=5/3$. The code accounts for metallicity dependent cooling by using the tabulated cooling functions of \citet{sutherlanddopita93} for gas temperatures of $10^{4-8.5}$~K, and rates from \citet{rosenbregman95} for cooling down to lower temperatures.
 
We have carried out two simulations of an isolated star forming galaxy: one where the high-mass single stars are initiated with the velocity kick from the velocity distribution given by \eqn{velocity_distribution} using $\beta=1.4$ (referred to as \kick), and a control simulation where high-mass single stars are not assigned any velocity kick (referred to as \nokick) as in traditional galaxy simulations.

The initial conditions are those of the isolated disc galaxy in the \agora project \citep{Kim+2014, Kim+2016}, set up to approximate a Milky Way-like galaxy following the methods described in \citet{Hernquist1993} and \citet{Springel2000}. Briefly, the dark matter halo has a concentration parameter $c=10$ and virial circular velocity $v_{\rm 200}=150\kms$. This translates into a halo virial mass $M_{\rm 200}=1.1\times 10^{12}\msun$ within $R_{\rm 200}=205\kpc$. The total baryonic disc mass is $M_{\rm disc}=4.5\times10^{10}\msun$ with $20\%$ in gas. The initial stellar and gaseous components follow exponential surface density profiles with scale lengths $r_{\rm d}=3.4\kpc$ and scale heights $h=0.1r_{\rm d}$. The bulge-to-disc mass ratio is $0.125$. The bulge mass profile is that of \cite{Hernquist1990} with scale-length $0.1r_{\rm d}$. The halo and stellar disc are represented by $10^6$ particles each, and the bulge consists of $10^5$ particles.

The galaxies are simulated in a box with sides $600\kpc$ with adaptive mesh refinement allowing for a minimum cell size of $9\pc$. The refinement strategy uses a quasi-Lagrangian approach which ensures a roughly constant number of particles ($\sim 8$) in each cell which reduces discreteness effects \citep{Romeo08}. Furthermore, the AMR scheme splits cells into 8 new cells, each with mass $m_{\rm refine}=4014\Msun$, when their sum of stellar and gas mass exceeds $8\times m_{\rm refine}$. Star formation uses a standard procedure with the star formation rate density in cells exceeding a density threshold, $\rho_{\star}$, given by
\begin{equation}\label{eq:star_formation}
    \dot{\rho}_{\star} = \epsilon_{\mathrm{SF}}\frac{\rho_{\mathrm{g}}}{t_{\mathrm{ff}}}\ \mathrm{for}\ \rho_{\mathrm{g}} > \rho_{\star},
\end{equation}
where $\epsilon_{\mathrm{SF}}$ is the star formation efficiency per free-fall time, $\rho_{\mathrm{g}}$ is the cell gas density and $t_{\rm ff}=\sqrt{3 \pi/32G\rho}$ is the local gas free-fall time \citep[see][for details]{Agertz+2013}. For this work we adopt $\rho_{\star}=100\cc$ and sample star particles with a mass resolution of $500\msun$, from which we remove mass for the HMS sampling at each star formation event. For the simulations considered here we use a constant $\epsilon_{\mathrm{SF}}=5\%$. 
Observationally, the efficiency per free-fall time averages $1\%$ in Milky Way giant molecular clouds (GMCs) \citep[][]{Krumholtz&Tan2007}, albeit with a large spread \citep[][]{Murray2011b,Lee2016}.

\section{Results}\label{sec:results}

\begin{figure*}
    \centering
    \includegraphics[width=\linewidth,trim={0 0 9.1cm 0},clip]{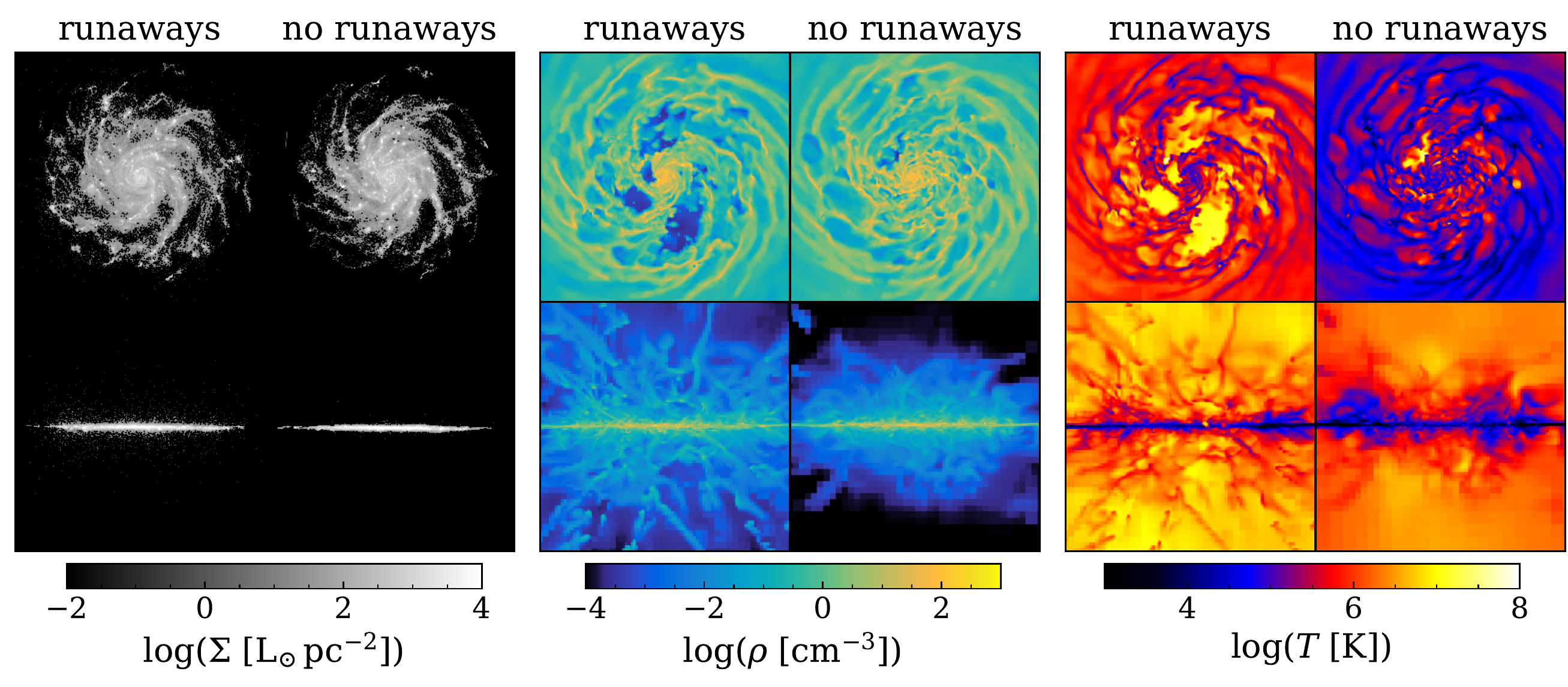}
    \caption{Projections of the stellar luminosity, gas density and temperature for face-on view (top row) and edge-on view (bottom row) from the last output of the simulations, i.e., after $250\Myr$ of evolution. The side of each panel is $24\kpc$. For each quantity we show the simulation taking runaway stars into consideration to the left, whereas the simulation ignoring these stars is shown to the right.}
    \label{fig:showcase}
\end{figure*}

We reemphasise that including runaway stars provides a means to relocate massive stars away from their dense formation sites before they end their lives in SNe, and our goal here is to explore how this process affects the evolution of a Milky Way-mass spiral galaxy. 

The two galaxies are evolved for $250\Myr$, a time during which they both are actively star forming. In the first $80\Myr$ the galaxies adjust to their initial conditions, and in the case of the \kick simulation there is a suppression in star formation after $120$ Myr (approximately one orbital time at the disc scale radius). After these $120\Myr$ both simulations has stable star formation rate. In this work all results taken as averages over time exclude this early phase unless otherwise stated. 

\fig{showcase} shows projections along the line of sight for luminosity, gas density and temperature for the galaxies at time $t=250\Myr$. Focusing first on the luminosity of the two galaxies, we find qualitative agreement between the two simulations, although we highlight that the \kick simulation has more stars in the inter-arm region as well as above and below the disc mid-plane. This indicates that the effect of runaway stars is subtle in the luminosity of the galaxy. However, the projected gas density and temperature largely differ between the two cases. Here runaway stars give rise to large ($\sim\!\kpc$), hot ($T\!>\!10^6$ K) bubbles in the inter-arm regions, as well as more prominent outflows (as seen on the edge-on views) compared to the \nokick model. In general, the gaseous halo is more structured with warm ($T<10^6$ K) clouds found far away from the disc ($\gtrsim 5\kpc$) in the \kick simulation.

\subsection{Star formation rates and ISM properties}\label{sec:SFR}
\begin{figure}
\begin{center}
	\includegraphics[width=0.95\columnwidth]{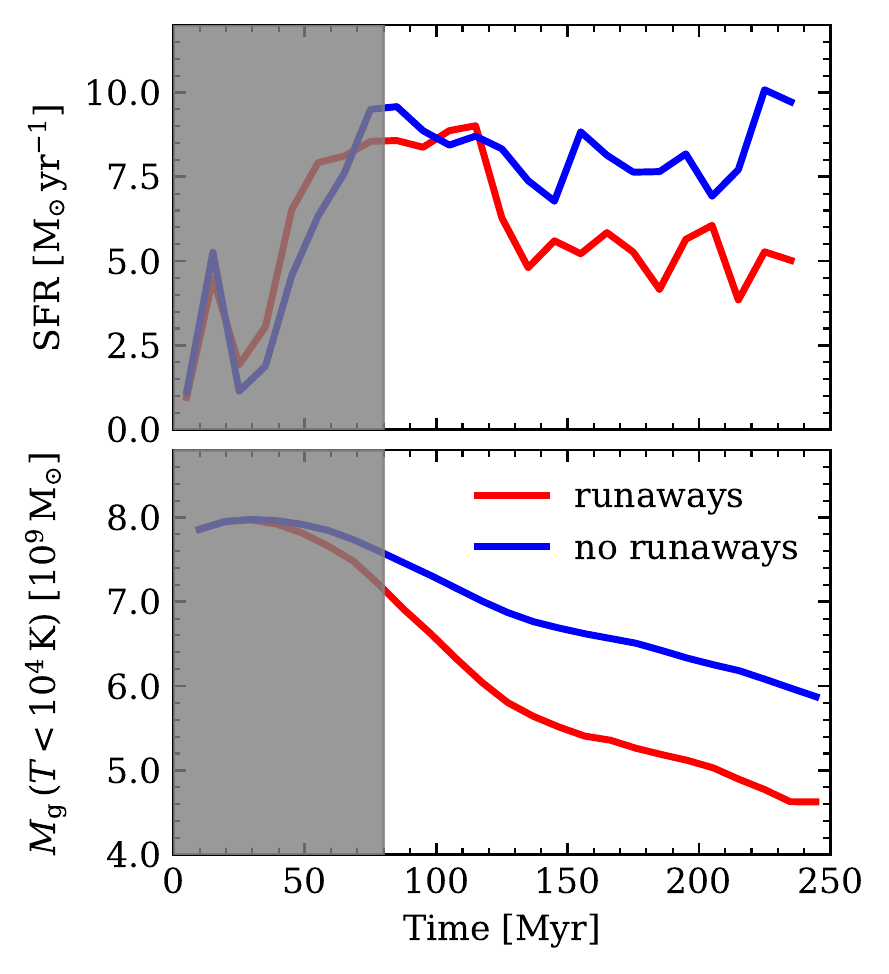}
	\end{center}
    \caption{{\it Top}: Star formation rate as function of time for the simulations with and without runaway stars. We neglect the transient caused by the simulation adjusting to initial conditions (grey region). Both simulations reach a steady state after approximately $120\Myr$, before which we see a quasi stable period with enhanced SFR (starting at $\approx80\Myr$). There is a suppression in star formation at $120\Myr$ in the model including runaway stars. {\it Bottom:} Total mass of the cold ($<\!10^4\,\mathrm{K}$) gas as function of time for the simulations, measured in a cylinder with a radii of $15\kpc$ and a thickness of $2\kpc$}
    \label{fig:disc_SFR}
\end{figure}

\begin{figure*}
    \centering
        \includegraphics[width=1.8\columnwidth]{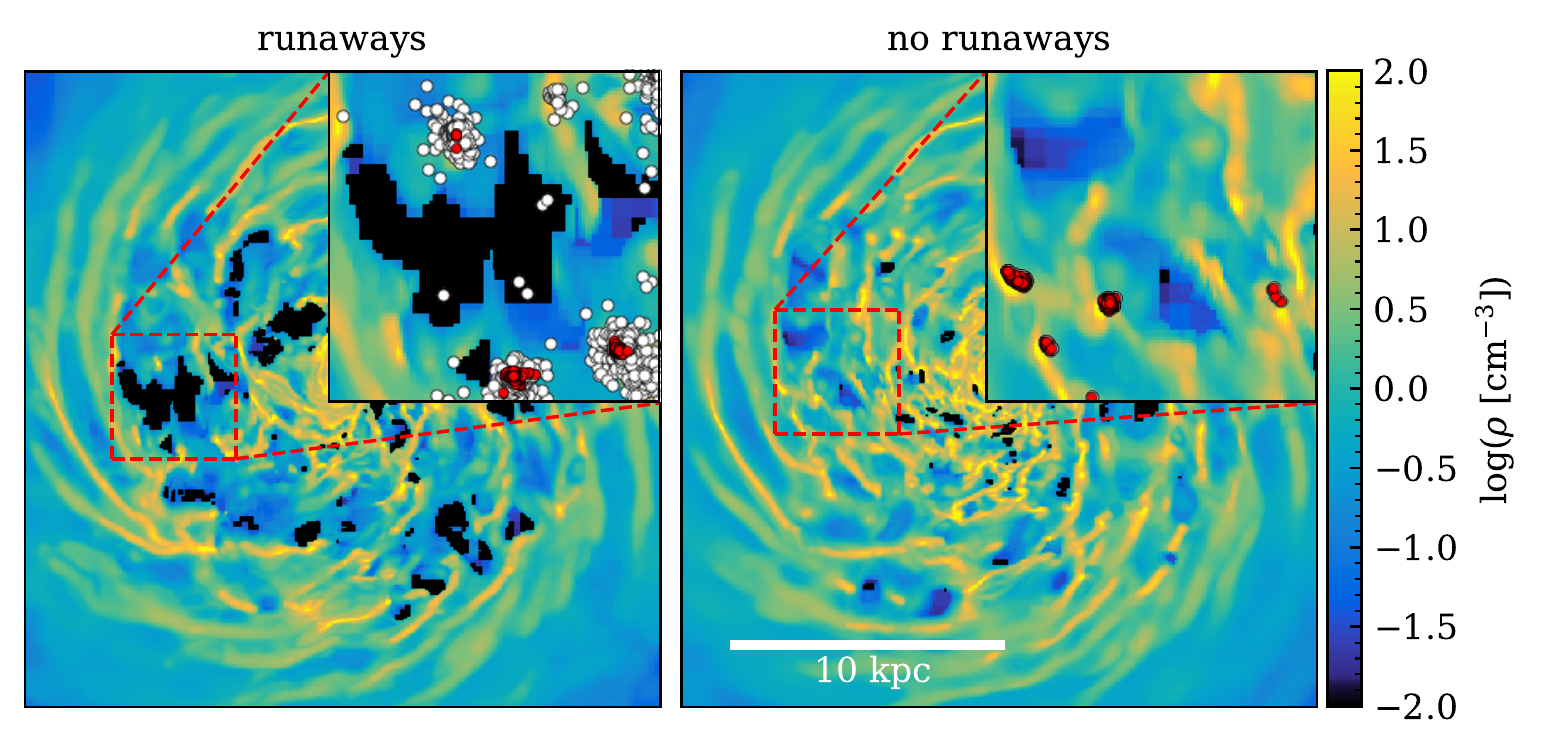}
    \caption{Projection plots of density for gas with temperature $<\!10^{4}\K$ for the model with (left) and without (right) runaway stars. The zoom in regions ($4.7\kpc$ wide) show the positions of young ($<10\Myr$, pre-SNe) star particles located within $50\pc$ of the mid-plane, with HMS particles in white and LMS particles in red. By construction, the two types exactly overlap in the \nokick case. In the \kick simulation a few HMS have traveled to regions completely void of $<\!10^{4}\K$ gas. These stars are likely to explode inside the bubble.}
    \label{fig:disc_density}
\end{figure*}

We next consider the star formation rate (SFR) of the two simulations, shown in the upper panel of \fig{disc_SFR}. Excluding the initial transient, both simulations have similar SFR ($8-9\Msunyr$) for the first $120\Myr$, followed by a decrease to a SFR of $5-6\Msunyr$ in the \kick simulation for the remaining time. This decrease is linked to the depletion of cold ($T<10^4$ K) gas, which is shown in the lower panel of \fig{disc_SFR}. The simulation with runaway stars features less mass in the cold phase at all times, which ultimately leads to a decline in the SFR\footnote{We note however that many other factors e.g. the level of gas turbulence, local star formation efficiency, GMC mass function will play a role in setting the global SFR.}.

During the period of high SFRs for both models ($80~{\rm Myr}<t<120$ Myr), more cold gas disappears in the \kick simulation compared to the \nokick simulation. Quantitatively, the total cold gas mass decreases at approximately $30\Msunyr$ and $15\Msunyr$ for \kick and \nokick respectively. Since the SFR is similar in both simulations at this epoch (thus having similar amounts of cold gas turning into stars), the \kick model must reduce the net gas mass more efficiently, either by removing it from this cold phase or inhibit warm gas from cooling more strongly compared to the \nokick model. To some extent both of these likely play a role. After $120\Myr$, the cold gas mass loss rate changes to $7-10\Msunyr$ in both simulations, with the \kick simulation now having less cold gas in total. For the \kick simulation this transition coincide with the decrease in SFR. This implies that even at low SFR, more cold gas is lost per unit of stellar mass formed in the \kick simulation.

\begin{figure*}
	\includegraphics[width=1.01\columnwidth]{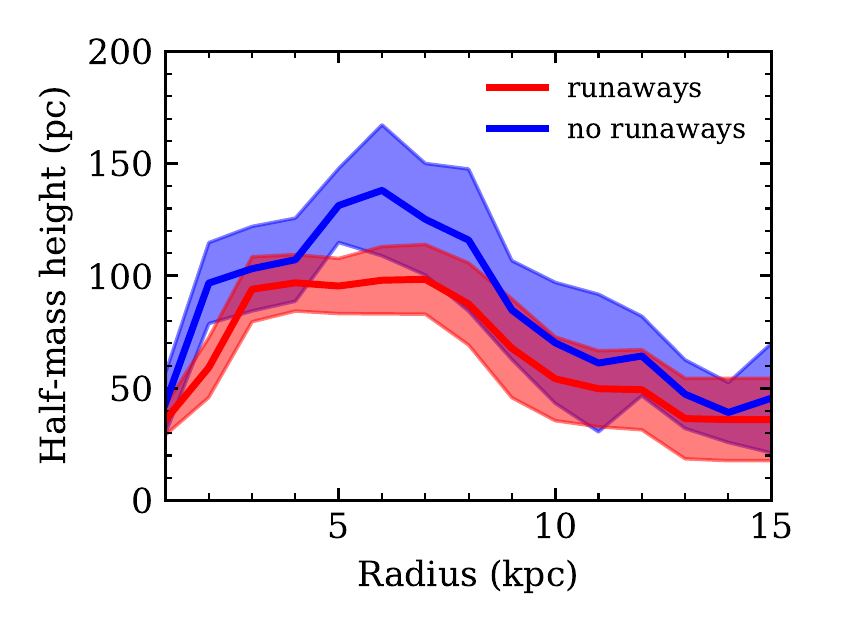}
	\includegraphics[width=0.99\columnwidth]{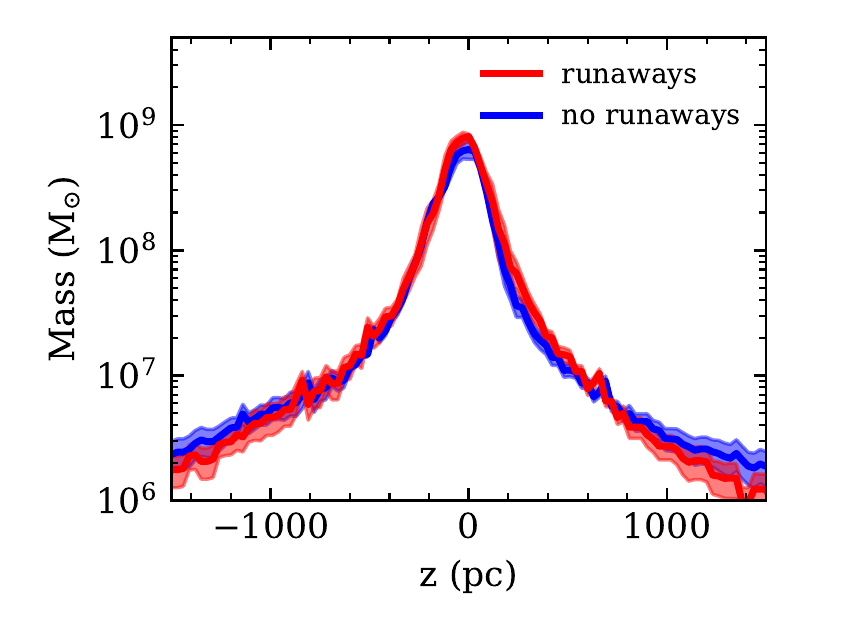}
    \caption{{\it Left:} Half-mass height of the cold ($<\!10^4\,\mathrm{K}$) gas mass as function of radius. Solid line shows the mean of all outputs and coloured areas show the standard deviation. {\it Right:} Vertical profile of the gas mass in the \kick simulation (red) and the \nokick simulation (blue). The filled line show the mean of all outputs and the coloured region is the standard deviation.}
    \label{fig:disc_structure}
\end{figure*}

To understand why the model with runaway stars predicts a more rapid removal of the cold ISM from the galaxy, we focus to the runaway mechanism, namely the effect of relocating SNe progenitors. \fig{disc_density} shows the projected gas density for the cold ISM ($T<\!10^4\K$). We selected outputs where both simulations included visible SN bubbles ($t=200\Myr$). The propensity of feedback from runaway stars to create large bubbles in the inter-arm regions is evident. For the \kick simulation there are young massive stars present in the voids, whereas for the \nokick all massive stars overlap with the low mass star particles by design. 

Initially, density contrasts are generated by stellar feedback and large scale gravitational instabilities, some of which evolves into bubbles. A crucial difference is that with the inclusion of runaway stars, the bubbles are resupplied with SN progenitors which keep injecting energy and momentum into them even though actual star formation is impossible in those regions. This repeated injection gives rise to $\kpc$-scale bubbles in the \kick model, and as we will discuss later it leads to energy venting into the CGM. Throughout the \kick simulation many such bubbles form, albeit with seemingly random timing.

Quantitatively, we find a clear excess of SNe explosions in gas with $10^{-5}\cc<\rho<10^{-3}\cc$ when including runaways, peaking at more than twice as many explosions at $\rho\sim 10^{-4}\cc$ with respect to the \nokick case. This implies a direct injection of energy in low density regions in contrast with the typical injection sites of SNe. This mechanism is visible in \fig{disc_density}, and the resulting heating is visible in the face-on projection of the gas temperature in \fig{showcase}. Finally, we note that the \nokick simulation features, not surprisingly, more SNe in gas with higher density compared to the \kick simulation.

Despite this seemingly violent impact on the cold ISM, the overall structure of the disc remains intact. To some degree, runaway stars alleviate the dense star forming gas from part of the feedback. This is demonstrated in \fig{disc_structure} were we show the structure of the cold gas in the disc. The left plot shows the vertical distance from the mid-plane of the disc within which half of the cold gas mass is located as function of radius. In the right plot we show the gas mass as function of vertical distance $z$. Although the introduction of runaway stars cause large bubbles in the inter-arm gas, the overall structure of the cold ISM is very similar in both simulations. In fact, runaway stars produce a slightly thinner disc. Furthermore, the vertical velocity dispersion of the cold gas is $\sigma_{\rm z}\sim20\kms$ at radius $R=5\kpc$ for both simulations, with a roughly linear decrease to $\sim5\kms$ at $10\kpc$ after which it flattens out. This is in close agreement with observed \hi kinematics in nearby spiral galaxies \citep[][]{Tamburro2009}. Hence, while runaway stars drive strong turbulence and creates large bubbles it does not strongly alter the morphology of the cold disc. This allows for the co-existence of thin galactic stellar discs and strong galactic outflows, an otherwise challenging aspect of galaxy models \citep[e.g.][]{Agertz+2011,Roskar+2014}.

In summary, runaway stars induce in more efficient coupling of feedback to diffuse gas. The more rapid depletion of the cold ISM found above is then a result of more vigorous galactic outflows which we quantify in the next section.

\subsection{Galactic outflows}\label{sec:outflows}

\begin{figure}
\begin{center}
	\includegraphics[width=0.95\columnwidth]{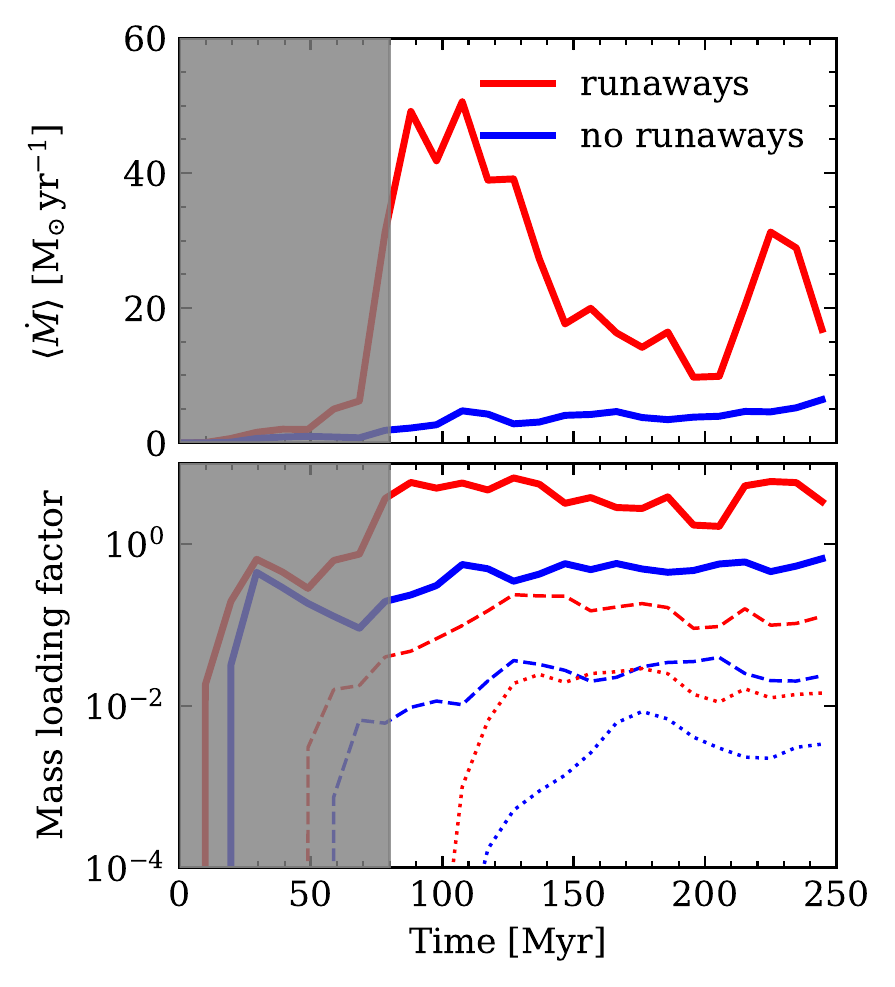}
	\end{center}
    \caption{{\it Top:} Vertical gas outflow rate as function of time for the two models, measured in cylinders of radius $12\kpc$ and thickness $2\kpc$, centred on $\pm5\kpc$ from the mid-plane. The grey region is neglected since the simulation is still adjusting to the initial conditions during this time. When measuring $\langle\dot{M}\rangle$ at different heights we see a similar time profile but re-scaled, i.e. both lines are shifted down by the same amount when measuring at larger height. The \kick simulation features epochs of very strong outflows due to the production of large bubbles, which keeps growing due to injection of supernovae energy by runaway stars. {\it Bottom:} Mass loading factor as function of time for the two models, computed by dividing the outflow rate with the SFR, which quantifies the efficiency of the stellar feedback in driving outflows. We show this for three different heights: $5\kpc$ with $2\kpc$ thickness, i.e., identical to top panel (thick lines); $50\kpc$ with $10\kpc$ thickness (thin dashed line); $100\kpc$ with $20\kpc$ thickness (think dotted line). The selected distances corresponds to roughly $0.025\,R_{200}$, $0.25\,R_{200}$ and $0.5\,R_{200}$.}
    \label{fig:disc_outflows}
\end{figure}

\begin{figure}
	\includegraphics[width=0.95\columnwidth]{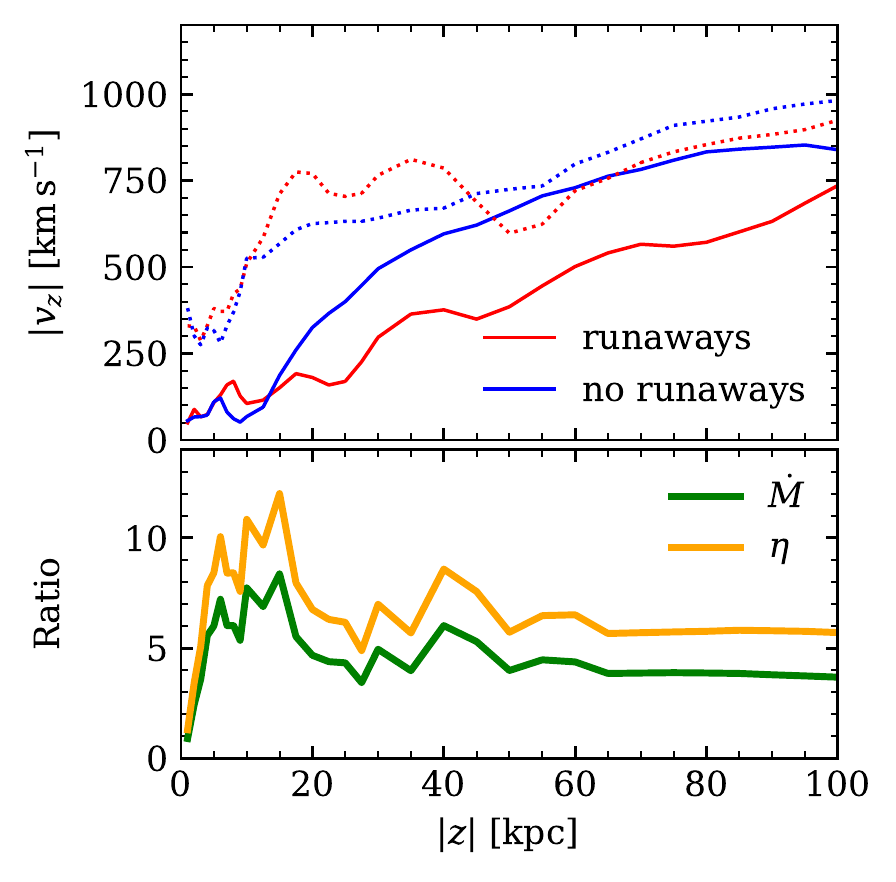}
    \caption{{\it Top:} Mean vertical velocity of the gas as function of vertical height, with warm gas ($2\times10^4<T<10^{6}\K$) as solid lines and hot gas ($10^6<T<10^{8}\K$) as dotted lines. {\it Bottom:} Ratio of outflow rates (green) and mass loading factor (orange) between the two simulations (\kick divided by \nokick) as function of vertical distance. To compute the outflow rates we used bin heights of $1\kpc$ for $|z|<10\kpc$, $2.5\kpc$ for $10\kpc<|z|<30\kpc$ and $5\kpc$ for $|z|>30\kpc$ in order to account for the decreasing resolution with increased vertical height. The lines shows the mean of all simulation outputs ignoring the first $80\Myr$.}
    \label{fig:disc_relative_prop}
\end{figure}

The top panel of \fig{disc_outflows} shows the vertical outflow rate as a function of time for both models. The rates are computed by summing cell-by-cell contributions in a cylindrical region with a radius of $12\kpc$ centred on the galaxy for $4\kpc<|z|<6\kpc$, considering only gas flowing in the vertical direction away from the mid-plane. The galaxy with runaway stars features significantly stronger outflows at all times, and right after the onset of the galactic wind ($t>80$ Myr) the mass loss rate is more than 10 times higher than without runaway stars. This strong outflows in the \kick simulation is likely linked to the reduction in cold gas mass, causing the reduced SFR. At $150\Myr$ this wind calms down, although it still removes mass at roughly twice the rate of the \nokick model. Furthermore, there are burst of outflows (the largest of which takes place at $200\Myr$) linked to runaway stars generating a number of large ($>\!\kpc$) bubbles (apparent in the middle panel in \fig{showcase}). 

In the bottom panel of \fig{disc_outflows} we present the evolution of the mass loading factor, defined as
\begin{equation}\label{eq:mass_loading}
    \eta = \frac{\langle\dot{M}\rangle}{\mathrm{SFR}}.
\end{equation}
Since $\eta$ is very sensitive to the height where it is measured, we show it for three different $|z|$ corresponding to roughly  $0.025$, $0.25$ and $0.5$ times the virial radius of the galaxy. For a given model, $\eta$ varies up to three orders of magnitude depending on where it is measured. However, the ratio between the simulations is mostly independent of where we measure it. We will return to this notion later as well in this section, as compare our models to observations in \sect{obscomp}.

Focusing on the innermost measurement, we find that at $100\Myr$, $\eta$ reaches its maximum at $\approx 0.5$ in the \nokick simulation. With the stronger outflows measured in the runaway model, combined with the reduced SFR, the mass loading factor is significantly higher throughout the entire duration of the simulation, with periods of $\eta$ reaching values of $\sim 5-10$. As we discuss further in \sect{longterm}, such an efficient wind driving implies a markedly different long term evolution for the galaxy.

Higher outflow rates do not necessarily imply faster galactic winds. In fact, in the \kick simulation we find that warm outflows ($2\times10^4<T<10^{6}\K$) carry more mass but travel at a lower velocity (see solid line in bottom panel of \fig{disc_relative_prop}) than in the \nokick simulation. If the galactic winds were purely momentum-driven, the outflow mass would depend on the wind velocity as $m\propto 1/v$. We measure the velocity to be less than twice as high in the \kick which then would imply an outflow mass no more than twice as high. However, the strong outflows in \kick yield $\sim\!3$ times more gas mass in the halo (as shown in \sect{structure}). Therefore, this feedback is not momentum-driven. Conversely, in the case of purely energy-driven winds, one would expect $m\propto 1/v^2$, i.e. at most four times more in the \kick, which is compatible with our mass measurements.

In both models, $v_z$ increases with vertical distance, indicating that outflowing gas is accelerated by the continuously shock heated hot phase. Moreover, in contrast to \nokick, the \kick simulation features epochs of vigorous winds, with bursts of fast moving, hot ($T\!>\!10^{6}\K$) gas originating from the large bubbles described earlier.

In the top panel of \fig{disc_relative_prop} we compare the mean outflow properties between the two models as function of vertical distance after $120\Myr$. Regardless of vertical distance from the disc, the model accounting for runaway stars always features an increased mass loss rate as well as mass loading factor. The impact of runaway stars on winds is hence not restricted to a narrowly defined region, but rather acts like a outflow strength `multiplier' all the way up to $\approx100\kpc$. If generic, we note that this effect could allow for a simple treatment of the effect of runaway stars in semi-analytical models of galaxy formation \citep[see e.g.][]{Somerville+2015} or cosmological simulations with phenomenological treatments of galactic winds \citep[e.g.][]{Vogelsberger2013}.

\subsection{Structure of gaseous halo}
\label{sec:structure}
\begin{figure}
    \centering
    \includegraphics[width=\columnwidth]{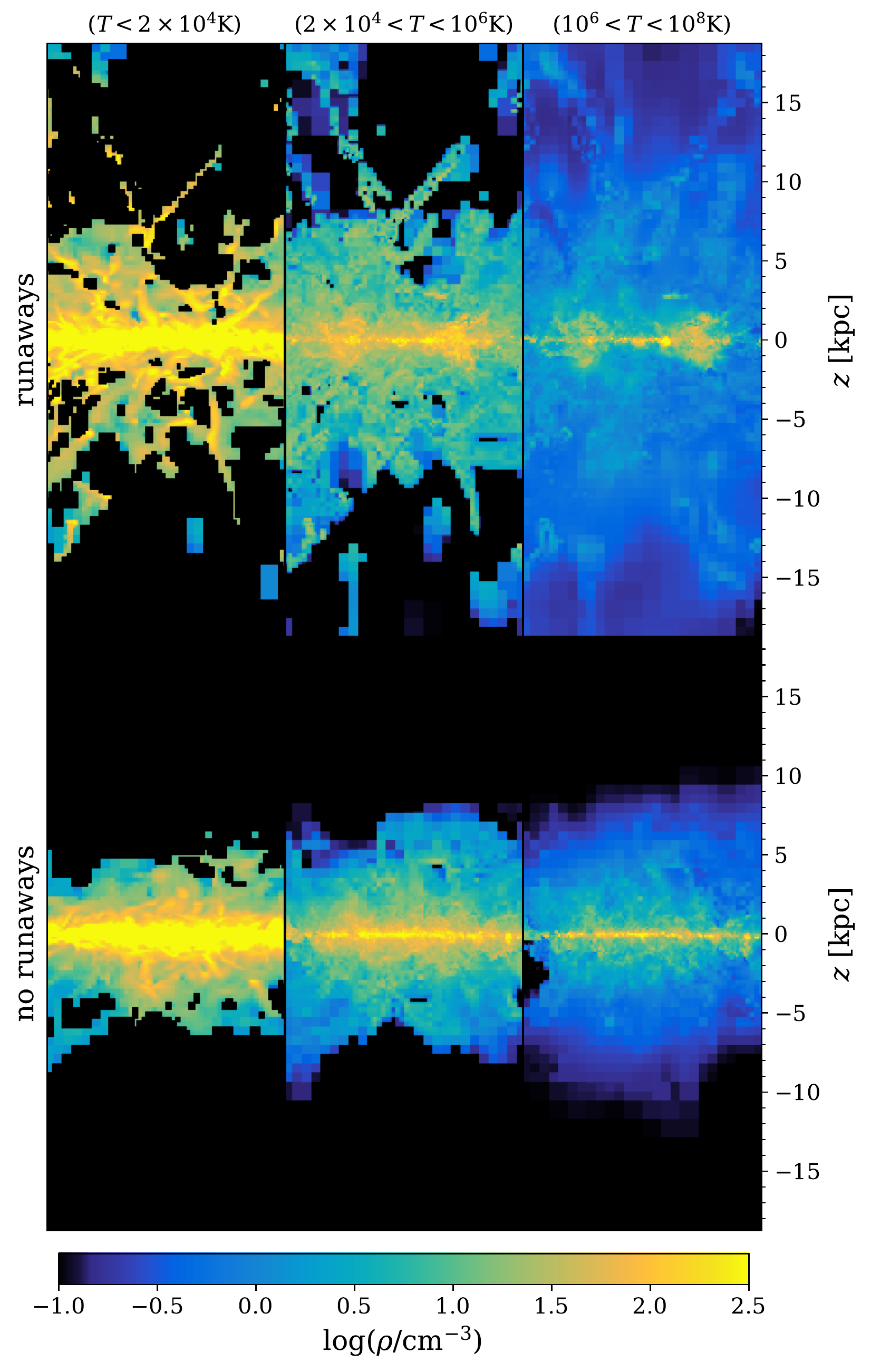}
    \caption{Projected density of gas in different temperature ranges (left panel: molecular gas; middle panel: warm ionised gas; right panel: hot gas) shown in panels with widths of $15\kpc$ in the edge-on view. In all three phases the outflows caused by the inclusion runaway stars provide a significantly larger vertical scale. When comparing the time evolution of the two simulations in this view, we find repeated large bubbles (most clearly seen here in the hot gas) in the \kick simulation. These are absent in the \nokick simulation. The snapshot shown here is the last output ($250\Myr$).}
    \label{fig:proj_rho_split}
\end{figure}

\begin{figure*}
	\includegraphics[width=1.6\columnwidth]{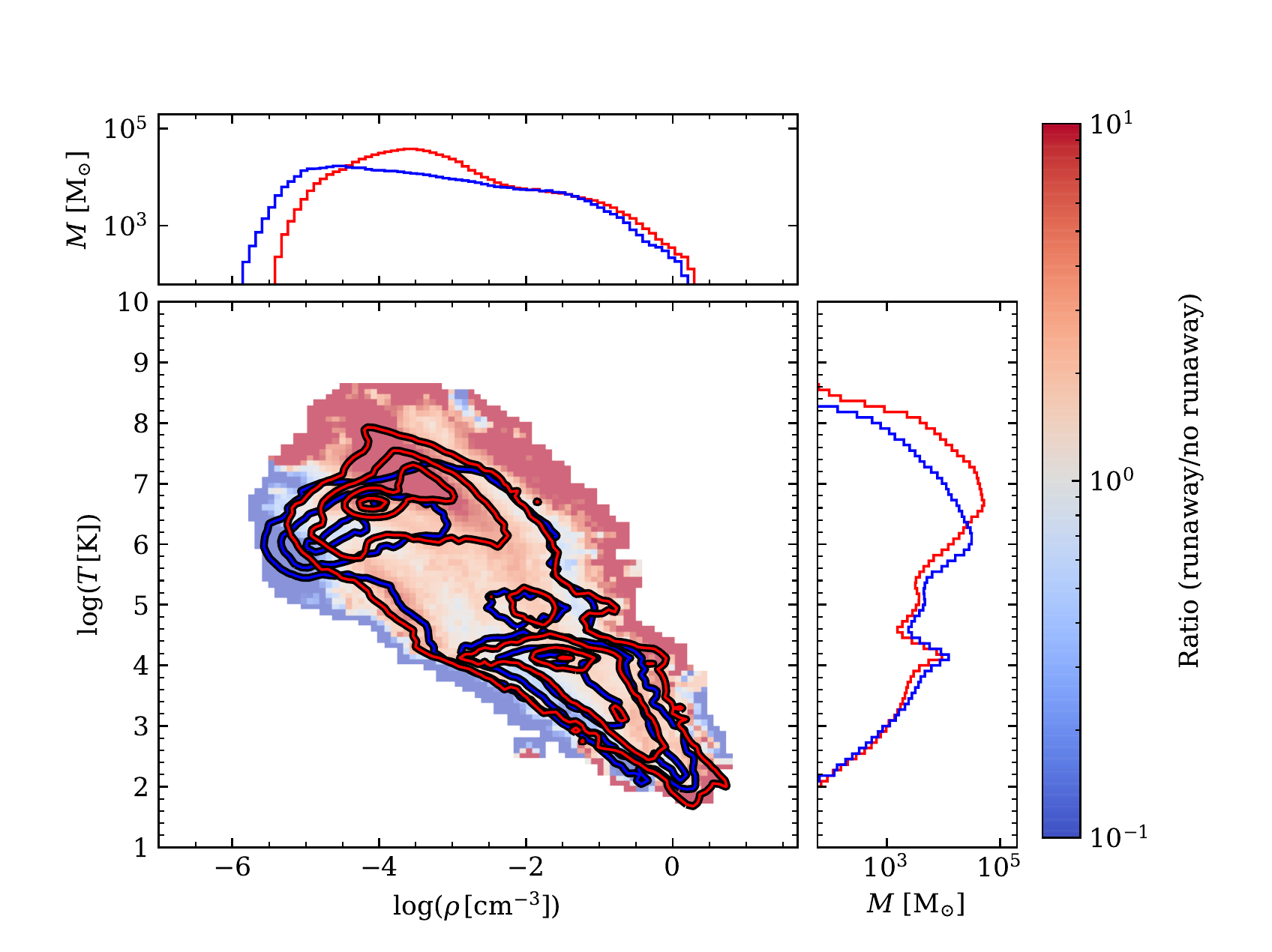}
    \caption{Comparison between the \kick (red) and \nokick (blue), as seen in the phase of the halo gas at the end of the simulation. The lines shows the mass distribution in gas phase, where contours are given for $1\%$, $10\%$, $40\%$ and $80\%$ and histograms shows the sum of all values along a given axis value. The background map shows the ratio between the models, coloured such that red indicated excess in \kick and blue indicated excess in \nokick.}
    \label{fig:halo_gas}
\end{figure*}

As already shown, runaway stars drive a strong galactic winds, launching large quantities of gas out in the halo. Despite the absence of cosmic inflows in our simulations and low resolution ($\sim100\pc$) in the halo, the circumgalactic medium (CGM) produced by the outflows does play a crucial role in the long term evolution of the galaxy, e.g. by determining the recycling of gas used for star formation. Unless otherwise stated, we focus now on gas within $40\kpc$ of the galaxy centre from which we remove the disc ($\pm2\kpc$ vertically and $15\kpc$ radially). 

\fig{proj_rho_split} shows the projected density shown for three different temperature ranges. Runaway stars produce a CGM which is more extended with the presence of long gas filaments and small clouds. The clouds rise from the disc and dissipate on timescales of tens of Myrs, in agreement with models of lifetimes of high velocity clouds around the Milky Way \citep[][]{HeitschPutman2009}. In our \kick simulation the motion of the clouds varies with some clouds being accreted back onto the disc, some buoyantly floating above the plane and some lifted outwards until dissipation. The filaments are the result of ram pressure exerted on these clouds by the hot medium. 

At $t=250\Myr$ the gas mass in the considered halo region is $9.0\times 10^{8}\msun$ and $2.9\times 10^{8}\msun$ in the \kick and \nokick models respectively. The excess mass in the \kick model is in very hot ($>10^{6}\K$), low density ($<10^{-2}\cc$) gas, shown in the contours of \fig{halo_gas} (as well as the histograms). The background colour of this plot gives the mass ratio between \kick and \nokick in different phases, where we see that the \kick model gives over one order of magnitude more gas mass in this phase, demonstrating the more efficient coupling of SNe energy to the halo. 

In the density histogram of \fig{halo_gas} we also see more mass for $\rho\approx1\cc$ in the \kick model. This is gas with temperature $\sim10^{4}\K$ which sits in clouds and the filamentary structures. Finally, we note that in the \nokick simulation there is more diffuse ($\rho\sim10^{-5}\cc$) gas in the range $10^{5}<T<10^{7}\K$ than in the \kick simulation. 

\section{Discussion}\label{sec:discussion}

\subsection{Implications for long term galactic evolution}
\label{sec:longterm}
We have demonstrated that simulations incorporating massive runaway stars feature galactic outflows with higher mass loading factors compared to models without. This leads to a more massive and structured galactic halo, and an overall reduction of the mass in the cold ISM. 

The star forming ISM is not only regulated more efficiently by galactic outflows, some fraction of the runaway stars travel far enough to deposit energy into the inter-arm regions and the inner galactic halo. This increases the thermal energy in diffuse gas, which in turn reduces the gas accretion rate of the galaxy and thus acts as a preventive feedback. Such a process is energetically more efficient than mechanically ejecting the same amount of gas from the ISM. This means that stellar feedback with runaway stars does not only lead to increased mass loading factors in galactic winds, but that it also includes a higher energy deposition into halo gas \citep{Li&Bryan2019}.

Although our simulations have only been carried out for 250 Myr due to the high computational cost associated with treatment of individual stars, the above mechanisms imply a more efficient regulation of star formation over $\Gyr$ timescales, and hence an overall impact on galaxy evolution over a Hubble time.  Applying our model in fully cosmological simulations, even for a limited amount of time, would be of great interest for shedding light on how runaway stars affect earlier phases of galaxy formation ($z>1$) when gas accretion rates are significantly higher than at the current epoch and efficient stellar feedback is know to be important \citep[e.g.][]{AgertzKravtsov2016}. Runway star physics may therefore aid in explaining the inefficiency of galaxy formation from e.g. abundance matching \citep[][]{Behroozi2019}, and will impact the precise galaxy mass range over which AGN feedback dominates stellar feedback \citep[]{Benson2003}.

\subsection{Comparison to other work}
\label{sec:prevwork}
The literature on galaxy scale simulations including runaway stars is limited due to the high computational cost of treating individual stars. Previous studies have therefore used alternative methods to overcome this challenge. Simulations of vertically stratified patches of discs have been used to compare clustered and random placement of SNe \citep[see e.g.][]{Korpi+1999,deAvillez2000,deAvillez&Breitschwerdt2004,deAvillez&Breitschwerdt2005,Joung&MacLow2006,Walch+2015,Martizzi+2016,Girichidis+2016,Gatto+2017}. Early works studied SNe placement based on density thresholds \citep{Korpi+1999} and showed that some SN explosions must occur in isolation in order to produce a realistic multi-phase ISM \citep[see, e.g.,][]{deAvillez2000,deAvillez&Breitschwerdt2004,Joung&MacLow2006,Walch+2015,Li+2017}. In agreement with our results, these authors have found an increased effect from stellar feedback as a result of randomly placed SNe. In fact, models with stellar feedback injected over a broad range of densities leads to a very turbulent ISM and an overall reduction of star formation. Additionally, \cite{Kim&Ostriker2018} found an increased mass loading factor as a result of including runaway stars, especially for hot gas, although this depends quantitatively on the vertical structure of the disc \citep{Li+2017}.

The volume filling factor of gas with different temperatures has, not surprisingly, been shown to be affected by randomly placed SNe \citep[see, e.g.,][]{deAvillez&Breitschwerdt2004,Walch+2015}. This is also the case in our work, with more of the volume filled by hot gas ($T>3\times10^5\K$) in the disc when including runaway stars (71\% vs. 54\%). We find that the opposite is true for the halo, where the dense clouds expelled by the strong winds somewhat reduce the volume filling factor in hot gas. 

For our model we account for the entire galactic disc, although this imposes constraints on the spatial resolution we can afford. As was pointed out by \citet{Martizzi+2016}, not accounting for the global geometry of the disc, as is the case for ISM patches, leads to unreliable wind properties due to incorrect treatment of the gravitational potential and the disc rotation. 

To our knowledge, no other simulation work of entire disc galaxies have included treatment of individual runaway stars, although alternative methods have been studied. A model by \cite{Ceverino&Klypin2009}, also used in simulations by \citet{Ceverino+2014} and \citet{Zolotov+2015}, includes the effect of runaway stars by applying velocity kicks to $10-30\%$ of all stellar particles (with masses $>10^4\msun$) representing entire stellar populations rather than individual massive stars. \citet{Ceverino+2014} showed that their model depend strongly on resolution with its effect disappearing at low-resolution ($60\pc$) compared to high-resolution ($14\pc$) simulations. This is not unexpected since the mean travel distance of OB runaway stars is of order $100\pc$ (see Section~\ref{sec:birth_kicks}) which sets a characteristic scale that needs to be resolved for these stars to differ from massive stars remaining in their natal regions. 

In order to alleviate the computational expense required by tracing the trajectory of individual stars, an intermediate step could be to distribute SNe explosions around their natal stellar population. \citet{Tress+2019} used this approach by injecting SNe energy at locations randomly sampled from a $5\pc$ Gaussian distribution. Although this captures some aspects like the spatial extent of star clusters, this is not representative of runaway stars which can travel significantly further and yield more complex distance distributions.

\subsection{Comparisons to observations}
\label{sec:obscomp}
Observational evidence for strong galactic winds is today ubiquitous e.g. in starburst galaxies \citep{Veilleux+2005,Rupke2018}. In this section we discuss how our simulations compare with the current observational data. We note that this comparison should be treated with caution since our simulations are idealised, isolated disc galaxies without their cosmological environment.

\begin{figure}
	\includegraphics[width=\columnwidth]{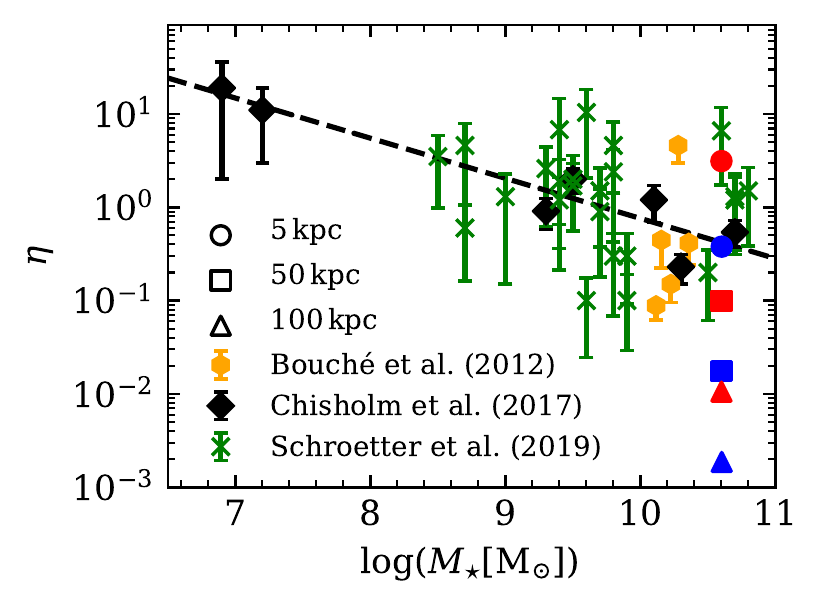}
    \caption{Mass loading factor as function of stellar mass comparing the results of our simulations to values observed. Our simulations, with \kick in red and \nokick in blue, are shown measured at three different vertical heights (see \fig{disc_outflows} for details). Orange markers show data for galaxies at redshift $\sim0.1$ from \citet{Bouche+2012}. Black diamonds shows data from nearby star-forming galaxies from \citet{Chrisholm+2017} with mass loading estimates computed from maximum outflow rates divided by star formation rates derived from UV spectra. The scaling relation derived in their work is shown as the dashed black line. The green crosses shows measurements from galaxies at redshift $\sim0.1$ by \citet{Schroetter+2019}. Although $\eta$ depends strongly on the where outflows are measured the ratio between the two simulations remain the same.}
    \label{fig:mass_loading_observations}
\end{figure}

Mass loading factors in external galaxies have been observed to depend on galaxy mass although there is a considerable spread with mass loading factors ranging between $\eta\sim 0.01-10$, even at a fixed stellar mass \citep{Martin1999,Bouche+2012,Chrisholm+2017,Schroetter+2019}. Mass loading factors are difficult to extract from observations due the incomplete census of various gas phases as well as their sensitivity to where in the galaxy one measures them. Since observations typically use quasars absorption spectra to study winds in external galaxies the random positioning of the quasar with respect to star forming galaxies makes this a challenging endeavour. In \fig{mass_loading_observations} we show a composite of observationally derived mass loading factors as function of stellar mass including a fit from \cite{Chrisholm+2017} together with our results. The data indicates a negative trend with stellar mass, as well as a large scatter. We show the final values of our results at three different heights above the disc plane. These values are in broad agreement with the entire spread in the observations, with the exception of the outermost measurements of the \nokick simulation. This implies that the spread in our simulations is due to either: 1) height above the disc plane; 2) the different stellar physics we consider, i.e., runaway stars. In the absence of information on the height of measurements (due to uncertain inclination of the disc) it is not possible to disentangle these two possibilities.

Other simulation work have modelled self-consistent driving of galactic outflows from stellar feedback in massive galaxies. Notably, \cite{Muratov+2015} analysed the FIRE simulation suite and found that for $M_\star\gtrsim 10^{10}\msun$, mass loading factors were consistent with $\eta\sim 0$ when measured at $0.25R_{\rm vir}\approx 50\kpc$ (see their figure 6). This is incompatible with the observations compiled in \fig{mass_loading_observations}, and illustrates that the role of stellar feedback driven winds in galaxies of this mass is not yet understood. The results of \citeauthor{Muratov+2015} may indicate a lack of additional physics such as runaway stars as our \nokick simulation also produce low, albeit non-zero, mass loading factors.

\begin{figure*}
	\includegraphics[width=2\columnwidth]{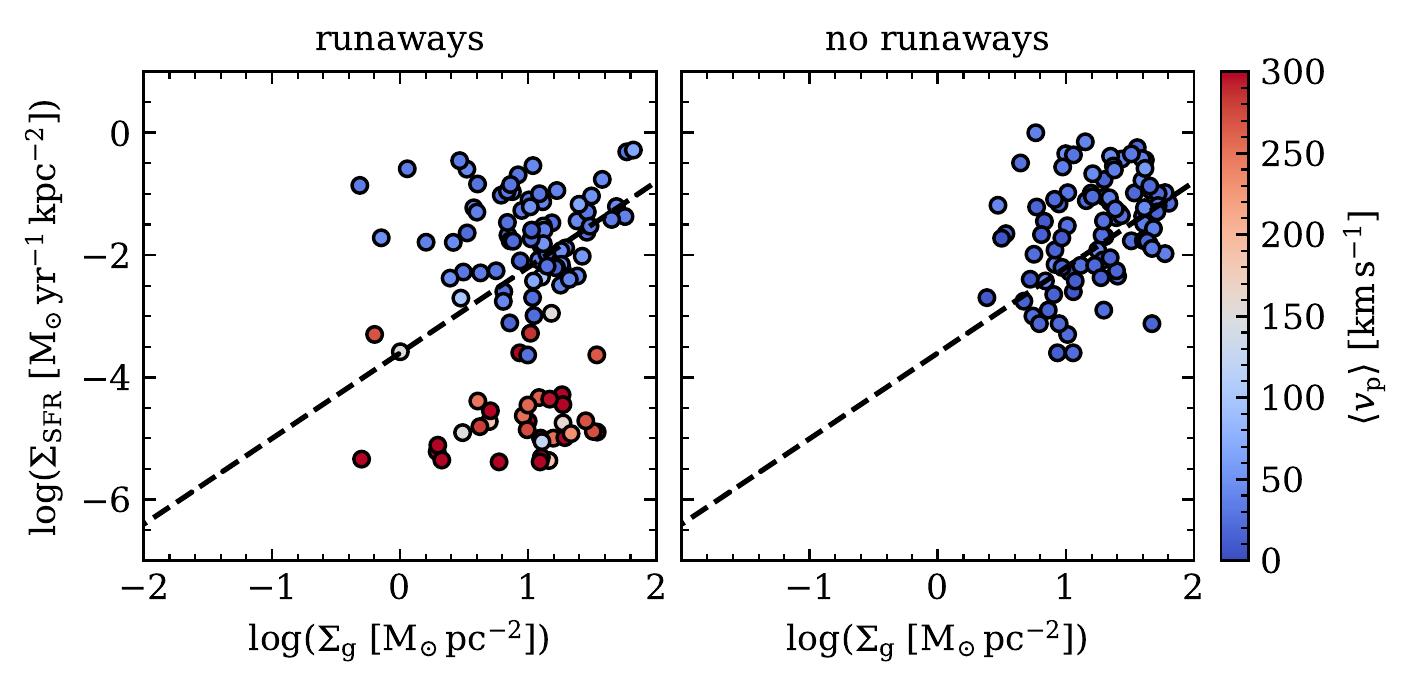}
    \caption{Resolved Kennicutt-Schmidt relation, $\Sigma_{\rm SFR}$ versus $\Sigma_{\rm g}$, comparing the runaway model (left) to the model ignoring runaway stars (right). Each point is a measurement from a $1\kpc$ square in the $24\kpc$ wide face-on view of the galaxy using data from our last output. To compute the star formation rate we use an age bin of $2\Myr$. The colours show the mean velocity of all these stars, calculated after removing the velocity component coming from the rotational velocity. The grey dashed line shows the empirical relation fitted by \citet{Daddi+2010} for disc galaxies. The \kick simulation includes a group of points with star formation rates $\sim 10^{-5}\Msunyr\,\mathrm{kpc}^{-2}$ which corresponds to runaway stars, as is indicated by the high velocity of the stars in these points.}
    \label{fig:resolved_KS}
\end{figure*}

Another striking effect of the runaway stars is the more structured and extended gas halo. The CGM is indeed observed to be highly structured \citep[e.g.][and references therein]{Werk2014}, with stellar feedback leaving a specific imprint on the properties of the circumgalactic gaseous haloes \citep[e.g.][]{Liang2016}. To date, no consensus exists on what is driving the fine scale structure of the CGM, partly due to the fact that galaxy scale simulations are far from resolving the (likely) parsec scale clouds that make up the halo on small scales (see discussion in Section \ref{sect:limitations}). We leave a study of the impact of run-away stars on halo absorber column density profiles and covering fractions for a future study.

In Section \ref{sec:structure} we noted that more halo gas clouds existed across a wide range of temperatures in the \kick simulation. They arise from gas lifted from the disc as well as halo gas compression from self-gravity followed by efficient cooling (indeed, cooling is most efficient at $T\sim {\rm few }\times 10^5$ K, with feedback driven perturbations leading to thermal instabilities \citet[][]{Joung2012} and cloud formation). In reality, these clouds may be destroyed via hydrodynamical instabilities as they fall back onto the disc \citep[][]{HeitschPutman2009}, a process that is likely not fully captured in our models. Regardless, these clouds could be interpreted as high velocity clouds (HVC), which are observed in the Milky Way halo \citep{Blitz+1999,Wakker2004,Moss+2013,Moss+2017}. The origin of the HVCs is still an open question \citep{Lockman+2019}, and although the most distant ones are likely from cosmic origin \citep[see e.g.,][]{Richter2012}, there are plenty of clouds observed within few tens of $\kpc$ \citep[see][]{Wakker+2008}, compatible with our \kick model. It is likely that at least some fraction of these clouds originate from outflows \citep{Quilis&Moore2001}, making the more efficient coupling of feedback energy to the galactic halo allowed by runaway stars an interesting addition to the formation scenario of HVCs.

Finally, we briefly highlight a curious effect found when including runaway stars in our simulations, namely their effect on the resolved Kennicutt-Schmidt (KS) relation. This relation is shown for the final output of both the \kick and \nokick simulation in \fig{resolved_KS}, together with the relation from \cite{Daddi+2010} (see also \citealt{Kennicutt1998}). With \kick we find regions of the disc which reside significantly below the canonical value, reaching $\Sigma_{\rm SFR}\sim 10^{-6}\Msunyr\kpc^{-2}$. Since our model has a finite mass resolution for star formation events ($\sim500\Msun$), a lower limit to star formation surface densities of $\Sigma_{\rm SFR}\approx10^{-4}\Msunyr\kpc^{-2}$, for the accounted time ($2\Myr$) and area ($1{\kpc}^2$), is expected. This is indeed the case for \nokick, since stars does not significantly change in mass over $2\Myr$. In contrast, in the \kick model this is enough for a fraction of runaway stars to venture away from their natal region and masquerade as inefficient star formation events. This is highlighted by the deviation in mean peculiar velocity of the stars shown by colour in  \fig{resolved_KS}. Runaway star may therefore provide an explanation for the extremely inefficient star formation observed in local spiral and dwarf galaxies \citep{Bigiel+2010} which we will focus on in forthcoming work (Andersson et al., in prep.)

\subsection{The impact of runaway stars in dwarf galaxies}
Dwarf galaxies have shallow potential wells and are expected to be strongly affected by stellar feedback, with higher mass loading factors than the cases studied here (see Figure~\ref{fig:mass_loading_observations}) resulting in low galaxy formation efficiencies \citep[][]{Behroozi2019}. 

To give us an indication of how runaway stars impact these low mass galaxies we set up two additional pairs of simulations. The first consists of a galaxy formed from collapsing gas in an analytical description of a dark matter halo. Its potential is set up according to a Navarro-Frenk-White (NFW) profile \citep{NFW1997} with $R_{200}=50\kpc$, where the rotational velocity is $35\kms$. This setup is identical to that of \citet{Teyssier2013}, except they used a ``live" halo described by particles. We adopt the same star formation and feedback recipes as is described in Section~\ref{sec:star-formation_and_feedback} and compared the evolution of one with and one without our runaway star model. After a settling phase the star formation and outflow rates were almost identical regardless of whether runaway stars are included or not. In addition to the isolated dwarf galaxy we simulated the formation of a ultra faint dwarf (UFD) galaxy ($M_{200}=10^9\msun$ at $z=0$) in cosmological setting. The setup used is the same as the fiducial simulation in \citet{AgertzEDGE2019}, except that we include our star formation and feedback models. Akin to the isolated dwarf galaxy we find a surprisingly small effect when including runaway stars, despite the mean travel distance for SNe progenitors being on the order of the half mass radius ($\sim300\pc$) of the simulated UFD. 

We attribute the lack of impact of runaways on dwarf galaxies, at least in terms of outflow properties, to the high porosity of the ISM which  arises due to the shallow potential. This implies that energy from any SN is likely to efficiently couple to diffuse gas within or outside galaxy, regardless of the inclusion of runaways. Nonetheless, since runaway stars can leave the galaxy entirely one may expect a different enrichment history for the CGM. How this affects the chemistry of the dwarf galaxy and its environment in the long term deserves a dedicated study. 

\subsection{Limitations of the model}
\label{sect:limitations}
Throughout this work we have outlined limitations to our model. In this section we summarise these points, and discuss possible improvements for the future. We highlight that this work is an exploratory study with the aim of giving a first approximation as to how runaway stars affect the physical processes in full galaxy simulations. 

Treating self-consistently the star-star interactions responsible for the velocity kicks is still not possible in galactic context. This is due to the extreme computational cost a star-by-star description would represent, and also because of the need of a direct collisional treatment of gravity, to resolve close encounters and binary evolution, which is not possible with galaxy simulation codes. Therefore, we must rely on sub-grid models to assign the kick velocities, and ignore possible variation with the environment. Nonetheless, future work should focus on how the results here depend on the velocity distribution, which in turn can be studied outside the galacic context \citep[see e.g.][]{Eldridge+2011,Perets+2012,Moyano+2013,Oh&Kroupa2016,Renzo+2019}
Furthermore, the results shown here assumes a runaway fraction of $14\%$ which is applied to all massive stars. Although this agrees with runaway fraction for O stars in a $10^{3.5}\Msun$ cluster \citep{Oh&Kroupa2016}, it is not universal and observational data indicates that this fraction can be significantly lower with variations with stellar type. \cite{Appelaniz+2018} estimated the fraction at $10-12\%$ for the O stars and $\sim6\%$ for B stars, which is also in agreement with those from \cite{Eldridge+2011} when correcting for completeness. The results shown here could therefore overestimate the impact of runaway stars. Future work should focus on how the results depend on changing there runaway parameters.

The resolution of our simulations is limited to $9\pc$. This is allows us to resolve the typical size of of the massive star forming gas clouds with approximately ten resolution elements, and this is essentially the scale runaway stars need to travel to reach low densities. Since this contrast in density is key to how the runaway stars affect the galactic evolution, an increase in resolution which allows for higher density contrast may also increasing the effect of runaway stars - in essence, less travel distance is required for more of the stars to reach low density gas to explode in. A suite of galaxy simulations with incrementally higher spatial resolution would be required to understand whether convergence can be reached.

Finally we note that due to the adopted adaptive refinement scheme the cooling length of the halo remains unresolved and we can only demonstrate systematic effects of the strong galactic wind caused by runaway stars. For a detailed discussion as to what this entails we refer to \citet{McCourt+2018,Hummels2019,vandeVoort+2019}.

\section{Conclusions}\label{sec:conclusions}
Using hydrodynamical simulations of Milky Way-like galaxies, we have investigated how their evolution is affected by the inclusion of runaway stars, a mechanism not commonly accounted for in galaxy simulations. Our model initialises massive stars as individual particles with velocity kicks randomly sampled from a typical distribution for natal star clusters. For our star-by-star treatment, we have implemented a new model for stellar feedback accounting for fast stellar winds and core-collapse SN. We compare the impact of this implementation to one with similar feedback physics but without any velocity kicks, thus ignoring runaway stars. In summary we find the following differences when including runaway stars:
\begin{enumerate}
    \item A significant fraction of SNe explode in low density gas ($\rho<10^{-3}\cc$), with more than a doubling of SN explosions at densities $~10^{-4}\cc$. By placing SNe progenitors in hot bubbles, where star formation is otherwise inhibited, the runaway mechanism significantly heats the inter-arm regions and drives strong galactic winds. This yields mass loading factors $\eta\sim 5$ when measured at the disc-halo interface -- an order of magnitude increase compared to models neglecting runaways. Both our models are compatible with observational values of $\eta$, which have large scatter due to uncertainty in where in the galaxy it is measured. Therefore, measurements of $\eta$ with accurate estimates of location are required to fully understand the importance of accounting for runaway stars.   
    \item In the halo we find three times more gas mass, primarily in the hot diffuse phase, as well as a population of dense clouds which dissipate on timescales of tens of $\Myr$s. This is mainly due to the large amount of gas lifted out by outflows. Furthermore, runaway stars are able to travel far enough to directly deposit SN energy into the halo, efficiently heating the CGM and preventing re-accretion of gas onto the disc. 
    \item The cold ISM ($T<10^4\K$) is less disturbed by stellar feedback since runaways can leave the star formation regions. This implies that, although runaway stars lead to increased feedback efficiency, there is little change to the overall structure of cold ISM. Therefore, the effect of runaway stars cannot be simply modelled by increasing stellar feedback, but an actual relocation of the feedback sources must be implemented. 
\end{enumerate}

One limitation of our model is the assumption of a universal velocity distribution for stellar birth environments. Contrary to this, properties such as cluster density, binary fraction and IMF variations ought to play a role. Models accounting for stellar-scale properties (e.g. binary stars) in cosmological models exist \citep[see e.g.,][]{Rosdahl+2018}, although without treatment for runaway stars, which requires tracing individual trajectories. A complete understanding of how runaways affects galaxy evolution might require a combination of these kind of models. However, treating even just a fraction of the stars as individual particles is very numerically expensive, therefore studying how our results depend on this velocity distribution is left to future work. Another limitation is that our model for stellar feedback ignores radiative processes. The strong luminosity of OB stars heats the surrounding medium before they explode as SN, thus affects how the energy is transferred to the gas. Adding radiative transfer to our already computationally heavy method is however beyond the scope of disc galaxy simulations. This would however be feasible for less massive galaxies, however, we do not see any significant effect of runaway stars for these galaxies. We attribute this to the high porosity of small galaxies implying that SNe location plays a minor role. As a final remark, the effect of runaway stars is clear in spiral galaxies for which the strong outflows lead to a more efficient regulation of star formation and in the long term this might impact the high end of the luminosity function.

\section*{Acknowledgements}
We thank Romain Teyssier for insightful discussions that improved our work. We thank the referee for helpful and constructive comments which greatly improved this work. The analysis of this work has made use of \texttt{yt} \citep{yt}, \texttt{pynbody} \citep{pynbody}, matplotlib \citep{matplotlib} as well as NumPy \citep{numpy}. We acknowledge support from the Knut and Alice Wallenberg Foundation. OA acknowledges support from the Swedish Research Council (grant 2014-5791). The simulations were performed on resources provided by the Swedish National Infrastructure for Computing (SNIC) at Lunarc. EA acknowledges financial support from the Royal Physiographic Society of Lund.  




\bibliographystyle{mnras}
\bibliography{ref}



\appendix



\bsp	
\label{lastpage}
\end{document}